# Chalcogenide Glass Planar MIR couplers for future chip based Bracewell Interferometers

H-D. Kenchington Goldsmith[*a], N. Cvetojevic[bcd], M. Ireland[e], P. Ma[a], P. Tuthill[d], B. Eggleton[c], J.S. Lawrence[bdf], S. Debbarma[a], B. Luther-Davies[a], S. J. Madden[a]

[a]Centre for Ultrahigh bandwidth Devices for Optical Systems (CUDOS), Laser Physics Centre, Research School of Physics and Engineering, Australian National University, ACT, 2601, Australia;
[b]The Australian Astronomical Observatory (AAO), Level 1, 105 Delhi Rd, North Ryde, NSW 1670, Australia;
[c]Centre for Ultrahigh bandwidth Devices for Optical Systems (CUDOS), School of Physics, University of Sydney, NSW 2006, Australia;
[d]Sydney Institute for Astronomy (SIfA), School of Physics, University of Sydney, NSW 2006, Australia;
[e]Research School of Astronomy & Astrophysics, Australian National University, ACT 2611, Australia;
[f]MQ Photonics Research Centre, Department of Physics and Engineering, Macquarie University, NSW 2109, Australia.

## ABSTRACT

Photonic integrated circuits are established as the technique of choice for a number of astronomical processing functions due to their compactness, high level of integration, low losses, and stability. Temperature control, mechanical vibration and acoustic noise become controllable for such a device enabling much more complex processing than can realistically be considered with bulk optics. To date the benefits have mainly been at wavelengths around 1550 nm but in the important Mid-Infrared region, standard photonic chips absorb light strongly. Chalcogenide glasses are well known for their transparency to beyond 10000 nm, and the first results from coupler devices intended for use in an interferometric nuller for exoplanetary observation in the Mid-Infrared L' band (3800-4200 nm) are presented here showing that suitable performance can be obtained both theoretically and experimentally for the first fabricated devices operating at 4000 nm.

[*]harry-dean.kenchingtongoldsmith@anu.edu.com

## 1. INTRODUCTION

The first exoplanets were discovered in the period from 1989 to 1991[1–3] and since then 1618 exoplanets have been listed in exoplanets.org as of 2016. The main driver for the increased rate of detection is the Kepler space telescope which uses the transit method of detection. The types of exoplanets that are favoured using this method are large Jupiter sized planets in close orbit around their host star. From space, with increased photometric precision the Kepler space telescope has also made some discoveries of smaller planets including Earth-like[4] and super-Earth[5] exoplanets.

To study exoplanetary formation inside a dust cloud or to take the spectra of exoplanetary thermal emission to determine atmospheric composition requires a direct imaging method in the Mid-Infrared (MIR). Recently direct imaging techniques have imaged planetary formation around a young star[6] and large exoplanets with large orbits from their host star[7,8] proving that this method is viable for exoplanetary research. One major step required for further studies is to extinguish the star light interferometrically, a so far non-trivial task due to the stability required and very limited range of available materials that can be readily formed into waveguides with suitable properties.

## 2. BACKGROUND

### 2.1 Interferometric Detection of Exoplanets

Rather than using one large telescope, long baseline infrared interferometry uses multiple telescopes to measure Fourier components of an image, enabling image reconstruction or model fitting. Such modelling may include a series of unresolved planets near a host star or simply the dust clouds around the central star. This is the method proposed in future exoplanet imaging space missions like TPF[9]. These missions seek to combine light such that exoplanets are distinct from their host star. One method of achieving this is to null the star light using interference. This method was proposed by Bracewell in 1978[10]. The idea is to take in light from a star and exoplanet from multiple telescopes then engineer the star light to destructively interfere whilst the exoplanet does not, due to its relative motion to the star.

Before the space missions launch, ground based interferometry is the only option, where the use of large monolithic mirrors, rather than interferometers, provide an excellent prototyping platform. Using a suitable Adaptive Optics system, atmospheric distortion can be effectively cancelled[11,12] to enable interferometry on sub-apertures of the telescope pupil. The light is sampled in the pupil plane with multiple non-redundant baselines using a microlens array. The *Dragonfly* project[13] is an example of such a scheme, which in addition, has a pupil remapper that maps the 2D baseline sample set to a 1D line array suitable for coupling to a photonic chip interferometer. The effectiveness of this approach is demonstrated in the NIR and now the challenge is to extend to the MIR.

### 2.2 MIR Photonic Chips

The standard silica based photonic chips used for integrated optics and successfully exploited in astronomy have low transparency beyond 2000 nm. It has been shown[14] that Chalcogenide glass (ChG) can be used to create low loss waveguides with minimal absorption features and low propagation losses across at least the 2000 - 5500 nm region and with the expectation that beyond 10000 nm the glass will remain transparent. The cited work showed that a ChG waveguide with a Germanium Arsenic Selenium ($Ge_{11.5}As_{24}Se_{64.5}$) core, refractive index of 2.609 at 4000 nm, fabricated on top of a layer of Germanium Arsenic Sulfur ($Ge_{11.5}As_{24}S_{64.5}$), refractive index 2.279 at 4000 nm, can transmit MIR wavelengths with ~0.2 dB/cm propagation losses. This waveguide, however, was essentially air clad leaving it vulnerable to atmospheric chemical and particulate contaminants and so unsuitable for interferometric operations. The device, in addition, used a rib waveguide which has intrinsic geometrical birefringence, similarly not ideal for interferometry. In this work, the prior demonstration was extended by fabricating a fully etched symmetric waveguide structure, Figure 1, to potentially eliminate birefringence and also cladding it with the same glass as the undercladding to symmetrise the optical mode and provide protection form atmospheric contaminants. The photonic chip used in this work was fabricated using thermal evaporation for layer deposition, and conventional positive resist photolithography and plasma etching for waveguide definition.

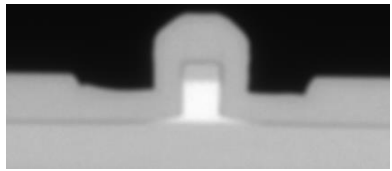

Figure 1. An optical image of the cross section of a nominally 2 μm x 2 μm waveguide with a core layer of $Ge_{11.5}As_{24}Se_{64.5}$, in white, and over and under cladding of $Ge_{11.5}As_{24}S_{64.5}$, in grey.

The waveguide design dimensions were chosen at 2 μm x 2 μm as this provides single mode operation in the design window, yielding a spot size of 2.6 μm full width at $1/e^2$ intensity. The minimum bend radius for such a waveguide design is 160 μm, enabling a compact circuit layout. Whilst the spot size is small and therefore incompatible with the low high numerical apertures in the pupil plane or from conventional MIR fibres, modelling confirms that uptapering the waveguide dimensions enables spot sizes of 10 μm to be efficiently coupled without exciting higher order modes. Importantly techniques have been demonstrated to enable such up-tapering with low loss[15].

This waveguide technology was then the basis for the demonstration of coupler devices for an initial Bracewell Interferometer design.

## 3. TOWARDS THE BACEWELL INTERFEROMETER – MMI DESIGN

From a purely practical view a Bracewell Interferometer extinguishes star light and leaves planetary light emerging with a usable contrast ratio for direct observation. Given multiple telescopes or pupil plane samples, the simplest means of achieving this is to combine the beams pair wise with appropriate phase control to destructively interfere the star light in the output port. Doing so requires optical elements that combine two or more beams to undertake the interference, plus appropriate phase control and potentially amplitude matching for maximum extinction. One such type of device is the so called multimode interference coupler (MMI also known as a Talbot device). This takes the concept of forming multiple images from a single point source[16] and confines it to a rectangular cavity. The 2x2 MMI has already been shown to outperform traditional evanescent coupling as a power splitter[17] in terms of the operational wavelength bandwidth where a given tolerance on the splitting ratio is imposed. MMIs also are more fabrication tolerant[17,18], and do not typically have process critical small elements that make them difficult to fabricate.

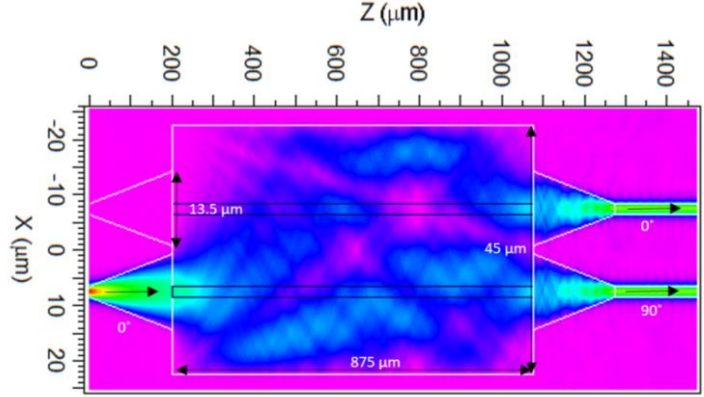

Figure 2. The simulated MMI from RSoft BeamProp of width 45 μm, length 875 μm with input and output tapers of 13.5 μm to 2 μm and a taper length of 200 μm. The black lines are purely to show which image is opposite and which is crossed with the initial image. The crossed image has the same phase as the input image and the opposite image (connected via the black line) is 90° out of phase.

A simulated MMI is shown as an example in Figure 3. The well documented formula for the position of an image (L), and the repetition of the image (M),[19–23] is shown in equation 1, where W is width, $\lambda$ is wavelength, $n_c$ is the refractive index of the cavity and N is the number of images that form at L, for M and N with no common factors.

$$L = \frac{M}{N} \frac{4 n_c \bullet W^2}{3\lambda} \qquad (1)$$

Looking specifically at the 2x2 MMI, when N = 2, a special case can be implemented to reduce the length of the cavity by a factor of 3, already taken into account in equation 1. This reduced case[23] requires the positions of the input waveguides to be at $\pm$ W/3 from the centre of the cavity, shown in Figure 3. In addition, to reduce the length of the MMI the first repetition, M = 1, is the only case considered. It is important to note the phase difference between the images in the output ports as this is vital for the Bracewell Interferometer. The image directly opposite the initial image, following the black outline in Figure 3, is 90° out of phase and the crossed image has the same phase as the initial image.

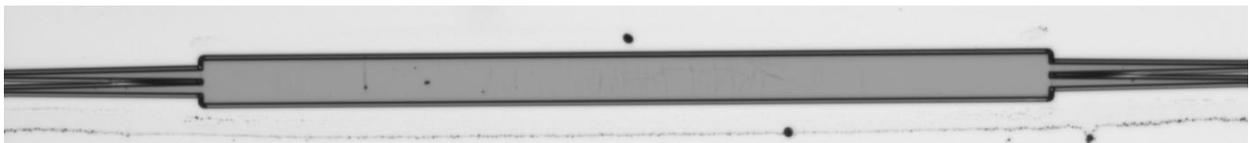

Figure 3. A top down optical image of a fabricated MMI with tapers of 8 μm to 2 μm.

By reciprocity, reversing the MMI in Figure 3 so that two coherent beams are input into the MMI provides a combined output from one port and a null in the other. This occurs as shown in Figure 3 only if the phase between inputs is 90°, for other phases in the 0° - 180° range, light will exit both ports following a sine squared relationship resulting in degraded interference. For this to be used as a Bracewell Interferometer the star light must have a 90˚ phase offset between inputs. The maximum contrast between star and planet light thus occurs at the star null port.

Note that for the simulated and measured MMIs in this paper a width of 45 μm, length of 845 μm and input ports that taper[18] from the 2 μm waveguide to 8 μm, as shown in Figure 3, were used. This has previously been shown to be a region with high optical performance yet good fabrication tolerances in simulations. The taper width of 8 μm is used rather than 13.5 μm to ease overcladding deposition during fabrication.

## 4. EXPERIMENTAL RESULTS

### 4.1 Experimental setup and waveguide performance

The light source used for testing was a custom built seeded optical parametric amplifier and a ND: $YVO_4$ picosecond pump laser with a wavelength of 1064 nm[24]. A CW tuneable semiconductor laser diode was mixed with the picosecond laser in a Lithium Niobate crystal to create the desired MIR wavelength covering the range of 3500 nm to 4200 nm. Schwarzschild designed reflective objectives with 36x magnification, 0.52 NA were used to inject and collect light from the waveguides. The OPA beam was focussed into the waveguide with an infinite conjugate objective and then the waveguide output reimaged using an objective with a 160 mm rear focal conjugate. Figure 4 (a) shows the focus of the first objective imaged at the finite conjugate with the second objective, i.e. it is the convolution of the two lens point spread functions. It displays the characteristic increased outer rings from the presence of an obscuration and the breaking of the rings into lobes caused by the support spiders for the small central mirror. The image was captured by a Xenics Onca InSb MIR camera which was also used to align the waveguide to a Gentec UM9B Pyroelectric Sensor. Figure 4 (b) and (c) shows the resulting images after coupled out from a 2 μm x 2 μm waveguide for both TE and TM polarizations of the incident beam.

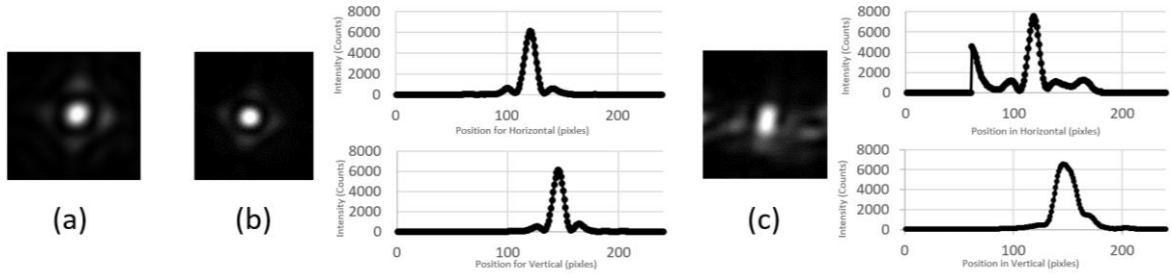

Figure 4. (a) Image input beam after passing through the two reflective objectives only. (b) TM polarization coupled into a 2 μm x 2 μm waveguide with the corresponding horizontal and vertical profiles centred at the maximum intensity. (c) TE polarization for the same waveguide with corresponding line profiles. Upper profile is horizontal cut lower is vertical cut. Pixel size is 30 μm and the lens has a magnification of 36x.

The expected spot in the output imaging plane would be similar to Figure 4 (a) but slightly smaller as the waveguide mode $1/e^2$ width of 2.6 μm is smaller than that of the focal spot from the lens (estimated at 4.8 μm assuming diffraction limited performance for the ~20% obscuration of the lens). This is only mirrored by the TM mode through the waveguide, whereas the TE mode appears to have an additional artefact. The line profiles in Figure 4 (c) compared to that of Figure 4 (b) give no indication that additional waveguide modes are coupling to the fundamental waveguide mode, the simulation of which is shown in Figure 5 (a). The additional peak in intensity appears to be caused by light coupling into a cladding mode. One such possible mode is shown in Figure 5 (b). The effective index (2.25) of this mode is below cut off and is therefore not a core mode, however; the core plus cladding acts as a waveguide in its own right, with air as a low index top cladding material and thermal oxide below, implying that this mode becomes guided. Future work will require a method of removing this pseudo-waveguide by planerising the overcladding.

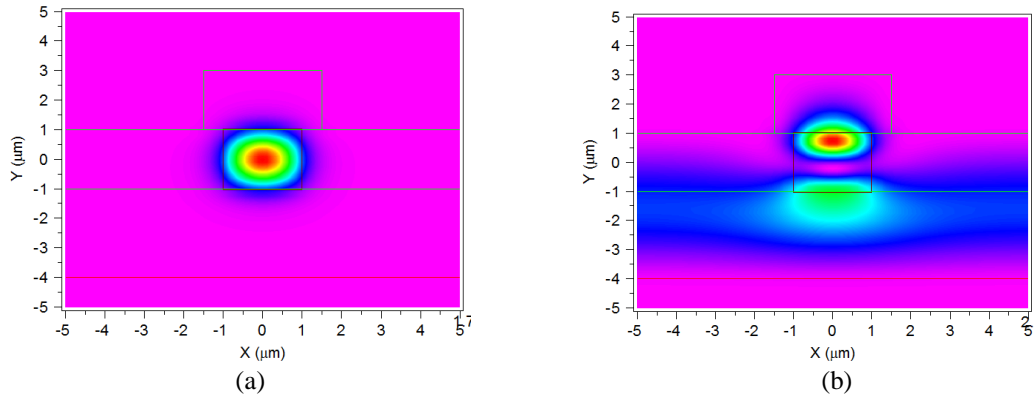

Figure 5. (a) The fundamental mode for TE, equivalent for TM, as simulated by the RSoft FemSim package. (b) A higher order TE mode that appears in the simulation of a 2 μm x 2 μm waveguide.

**4.2 Multimode Interference Coupler Performance**

The critical performance parameter in a Bracewell Interferometer, for the MMI, is the splitting ratio across the operational bandwidth, as this determines the depth of the null - the extinction. Based on modelling performance for the fabricated MMI shown in in Figure 3 an MMI design as per Section 3 provides a theoretical split of 50% across the 3800 - 4200 nm operating window. Measurement of the device shown in Figure 3, in the "clean" TM mode, produced the data shown in Figure 6 for light launched alternately into each of the two input ports.

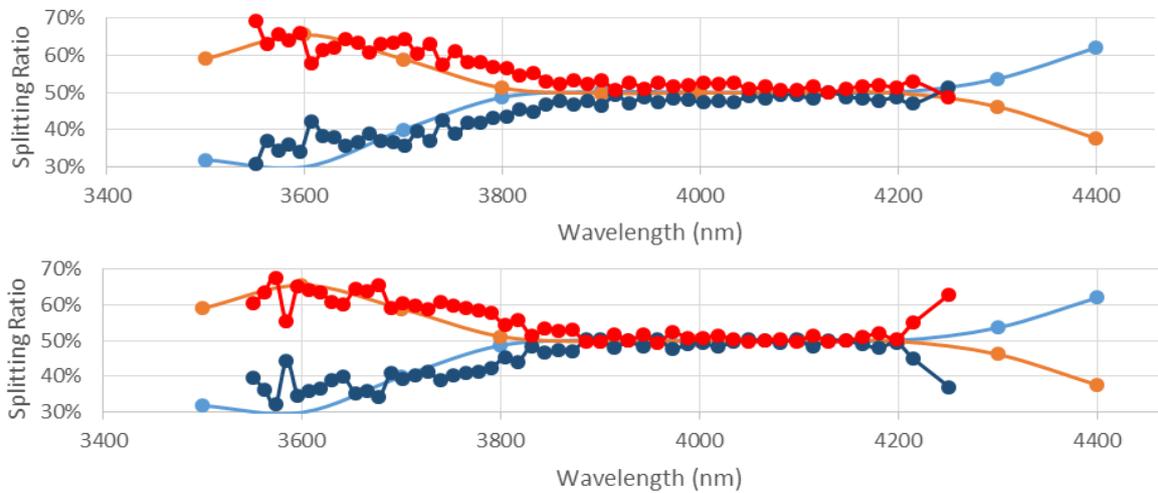

Figure 6. MMI as shown in Figure 3 with incident light through the right (top) and then left (bottom) input port. The blue and orange lines are the theoretical spits of the launched power through the MMI based on Rsoft BeamPROP full vector simulations using a unique mode computed for each wavelength independently. The red and dark-blue lines are the experimental results.

Here the measured performance is quite a good match for the predicted performance though the coupling ratio deteriorates relative to the predicted performance at the short end of the operational window. Further work is required to understand the source of this deviation, but it is clear that the devices operates largely as designed therefore enabling the construction of multiport nullers in the MIR. Future works will also need to use an additional CW laser to seed the picosecond OPA to measure at longer wavelengths and confirm the splitting ratio behaviour at longer wavelengths to help match theory with experiment at a larger bandwidth. In addition, the detection noise visible in the traces requires more investigation so that the reliability of the results can be confirmed.

## 5. CONCLUSION

Chalcogenide waveguide based MMIs have been shown experimentally for the first time to be suitable devices to build interferometric nullers in the MIR based on the principles of the Bracewell Interferometer. The astronomical L' band has been investigated as a first test with the broad band splitting demonstrated to be close to the theoretically predicted performance. To improve this MMI, further work is required in reducing the impact of a parasitic cladding mode that is excited by the TE fundamental mode, thus, creating a polarization independent device. The loss of the system needs further investigation in order to state without any doubt that this device is the best device for astrophysical applications in the MIR.

## ACKNOWLEDGMENTS

This research was supported by the Australian Research Council (ARC) Centre of excellence for Ultrahigh bandwidth Devices for Optical Systems (CUDOS - CE110001018). We also acknowledge the assistance provided though facilities usage and staff support in the ANU node of The Australian National Fabrication Facilities.